\documentstyle[aps,pre,epsfig,floats]{revtex}
\begin{document}
\twocolumn[\hsize\textwidth\columnwidth\hsize\csname@twocolumnfalse%
\endcsname
\title{Tests of Dynamical Scaling in 3-D
Spinodal Decomposition}
\author{S. I. Jury$^1$, P. Bladon$^1$, S. Krishna$^{1,2}$ and M. E.
Cates$^1$}
\address{$^1$Department of Physics and Astronomy, University of
Edinburgh\\ JCMB King's Buildings, Mayfield Road, Edinburgh EH9 3JZ,
GB}
\address{$^2$ Proctor Department of Food Sciences, University of Leeds, 
Leeds LS2 9JT, GB}
\maketitle
\date{\today}
\begin{abstract}
We simulate late-stage coarsening of a 3-D symmetric binary
fluid. With reduced units $l,t$ (with scales set by viscosity, density
and surface tension) our data extends two decades in $t$ beyond
earlier work. Across at least four decades, our own and others'
individual datasets ($< 1$ decade each) show viscous hydrodynamic
scaling ($l\sim a + b t$), but $b$ is {\em not} constant between runs
as this scaling demands. This betrays either the unexpected intrusion
of a discretization (or molecular) lengthscale, or an exceptionally
slow crossover between viscous and inertial regimes.
\end{abstract}
\vspace{0.3cm}
PACS numbers: 64.75+g, 07.05.Tp, 82.20.Wt 
\bigskip
]
When an incompressible binary fluid mixture is quenched far below
its spinodal temperature, it will phase separate into domains of
different composition. For symmetric (or nearly symmetric) mixtures,
these domains will, at late times, form a bicontinuous structure, with
sharp, well-developed interfaces. The late-time evolution of this
structure in three dimensions remains incompletely understood despite
theoretical \cite{siggia,furukawa,bray} experimental
\cite{experiments} and simulation \cite{alexander,laradji,lebowitz,appert}
work over recent years.

In the present work, we use the DPD (dissipative particle dynamics)
simulation algorithm \cite{dpd} to access length and time-scales far
beyond those reached previously. (Details of the simulations will
appear elsewhere \cite{jury}.) When combined with other datasets
\cite{alexander,laradji,lebowitz} our results allow a severe test of
the dynamical scaling ideas which underlie most theoretical treatments
\cite{siggia,furukawa,bray} and data analyses
\cite{experiments}. We conclude that dynamical scaling is in doubt,
perhaps due to the intrusion of a molecular lengthscale through the
physics of topological reconnection events. An alternative explanation
of the results, based on a universal but extremely slow crossover, is
also carefully examined.

As emphasized by Siggia \cite{siggia}, the physics of spinodal
decomposition involves capillary forces, viscous dissipation, and
fluid inertia. Indeed, assuming that {\em no other} physics enters,
then the parameters governing the behavior are the interfacial tension
$\sigma$, fluid mass density $\rho$, and viscosity $\eta$. (We now
specialize to 50/50 mixtures with complete symmetry of the two
species. Any asymmetries in composition, thermodynamics or viscosity
\cite{novik} provide additional control parameters.) From these three
parameters can be constructed only one length, $L_0 =
\eta^2/\rho\sigma$ and one time
$T_0 = \eta^3/\rho\sigma^2$. We now define the lengthscale $L(T)$ of
the domain structure at time $T$ via the structure factor $S(k)$ as
\cite{footlength} $L = (2\pi) \left( \int k S(k) dk /\int S(k) dk 
\right)^{-1}$. The 
exclusion of other physics in late-stage growth then leads us to the
dynamical scaling hypothesis
\cite{siggia,furukawa}:
\begin{equation}
l = a + f(t) \label{dynscal}
\end{equation}
where we define reduced time and length variables via $l \equiv L/L_0$
and $t \equiv T/T_0$.  Since dynamical scaling should hold only after
interfaces have become sharp, and transport by molecular diffusion
suppressed, we have allowed for a nonuniversal offset $a$ in
eq.\ref{dynscal}. Thereafter the scaling function $f(t)$ should
approach a universal form, the same for all (fully symmetric,
deep-quenched, incompressible) binary fluid mixtures.

It was argued further by Siggia \cite{siggia} that, for small enough
$t$, fluid inertia is negligible compared to viscosity, whereas for large
enough $t$ the reverse is true. This imparts the following asymptotes
to the function $f$:
\begin{eqnarray}
f & \to & b t  \;\;\;\;\;\;\;;\;\;\; t\ll t^* \label{viscous}\\
f & \to & c t^{2/3}  \;\;\;;\;\;\; t\gg t^* \label{inertial}
\end{eqnarray}
where, if dynamical scaling holds, the amplitudes $b$ and $c$ must be
universal, as must the crossover time $t^*$ (defined {\em e.g.} by the
intersection of asymptotes on a log-log plot).  Note that the Reynolds
number Re $= (\rho L/\eta) dL/dT = f \dot f$ which becomes large in
the inertial regime, eq.\ref{inertial}.

Perfectly symmetrical fluid pairs do not exist in the laboratory, but
computer simulations allow us to test the validity of eq.1, on which
the wider interpretation of experiments crucially depends
\cite{experiments}.  In the viscous regime (eq.\ref{viscous}), the
scaling reduces to $L(T) = A + B T$ where $A$ is nonuniversal and $B =
b\sigma/\eta$. This linear law has been reported by several
groups \cite{koga,puri,alexander} (see also
\cite{ma,appert,shinozaki}) but only in two recent cases
\cite{laradji,lebowitz} were reliable $\sigma$ and $\eta$ values
obtained, as are needed to find $b$. In both of these, the offset
$A$ was significant, and the linear regime (straight part of the
curve at late times) spanned much less than a decade.  In reduced
units
\cite{footlength}, we find that the data of Ref. \cite{lebowitz}
describes times in the range
$1\le t
\le 3$ with a value of $b = 0.3$. However the MD data of
Laradji et al \cite{laradji} has $60 \le t \le 140$ and $b =
0.13$.

The discrepancy over $b$ (see also \cite{alexander}) cannot simply be
brushed aside. For if dynamical scaling (eq.\ref{dynscal}) applies,
and both simulations
\cite{laradji,lebowitz} are (as claimed) in the viscous
regime (eq.\ref{viscous}), then these two $b$ values should both be
the same \cite{footcomp}.  It is thus premature to conclude that any
universal regime of viscous hydrodynamic scaling (eq.\ref{viscous})
had yet been observed in computer simulations.

To clarify this important issue, we have conducted several new
simulations which vastly extend the range of timescales explored: we
probe $750 \le t \le 45,000$. This was done using the DPD algorithm
which combines soft interparticle repulsions with pairwise damping of
interparticle velocities and pairwise random forces \cite{dpd}. The
latter conserve momentum, leading to a faithful simulation of the
isothermal (and, in this study, effectively incompressible
\cite{footcomp}) Navier Stokes equation at large lengthscales. Among
several advantages of DPD over MD, exploited below, is that the
viscosity of a DPD fluid can be varied {\em independently} of its
thermodynamics.

To describe our DPD parameters, we briefly switch from reduced
physical units ($l,t$) to ``DPD units": the range of the repulsive
interaction is unity, as is the particle mass.  We further set
$k_B{\cal T}=1$ (${\cal T}$ is temperature).  With the form of
repulsion used by Groot and Warren \cite{dpd}, we chose a particle
density 10 and energy parameters $a_{11}=a_{22} = 20, a_{12} = 100$,
which is a deep quench (${\cal T}_c/{\cal T} \simeq 80$)
\cite{jury}. The timestep was $0.01$ \cite{dpd}, giving measured $\cal
T$'s within 2 percent of the nominal value.  Integrating the
microscopic stress across a flat fluid-fluid interface
\cite{allenrowlwidom}, the interfacial tension was found as $50.6\pm
0.2$. For each damping parameter $\gamma$ the viscosity was found from
the the mean stress in steady shear under Lees-Edwards boundary
conditions \cite{lees}; values varied between $\eta = 2.6 \pm 0.2$
($\gamma = 1$) and $\eta = 12.2 \pm 0.5$ ($\gamma = 30$) \cite{jury}.

Most runs were performed on a 512 node Cray T3D, with a typical run
time of several thousand processor-hours. Resources allowed one
or two full-sized runs for each viscosity.  The simulation
box for these contained $10^6$ particles with periodic boundary
conditions. Thorough tests of scaling and data collapse for $S(k)$
were made \cite{jury}. Finite size effects became apparent when the
structural lengthscale $L$ exceeded about half the box size ($L \ge
\Lambda/2 \simeq 20$); data beyond this was excluded from our fits
for
$f(t)$. We also excluded an ``early stage" portion of each run;
this was judged by eye from the shape of the $L(T)$ plot. (Possible
resulting bias is considered below; little would be changed had we
instead applied a sharpness criterion to the observed interfaces.)

The datasets for $L(T)$ (DPD units) are presented in Fig.1
(inset). Excluded early time data is shown dotted, as is some data for
$L \ge \Lambda/2$. Slight wobbles in the fitted parts of the curves
represent sampling errors in $L$ arising because $L/\Lambda$ is not
small; these vary between duplicate runs and appear distinct from the
direct finite size (saturation) effects arising for $L \ge \Lambda/2$
\cite{footbased,footfinite}. Fig.1 shows the same data (with offsets 
subtracted) on a log-log plot.
\begin{figure}
\begin{center}
\leavevmode
\epsfxsize=0.35\textwidth
\epsfbox{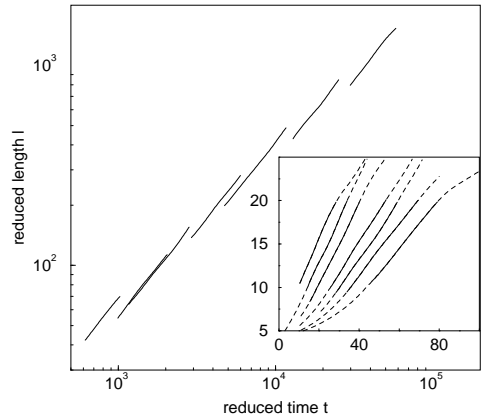}
\caption{Inset: Raw DPD data; $L$ vs. $T$ for viscosities (left to right) $
\eta = 2.6, 3.5, 4.6, 6.2, 8.2, 9.8, 12.2$. The datasets for $\eta =
6.2 , 9.8$ are averages of two runs. Main figure: the same data in
reduced units (log-log) with offsets (found by linear extrapolation to
$t=0$) removed.}
\end{center}
\end{figure}
Note first that, since most of the plots in Fig.1 show upward
curvature at early times, our elimination of early time data will bias
{\em downward} any estimate of the quantity $z=d\ln f /d \ln t$ (a
true or effective scaling exponent). Despite this, only for the
smallest viscosity run (if that) is there appreciable direct evidence
for an exponent $z <1$ as predicted for $t \gg t^*$
(eq.\ref{inertial}). A fit to eq.\ref{dynscal} with $f = c t^z$ in
fact gives $z = 0.88$ whereas all but one of the other viscosities
give $1.10 \le z\le 1.17$ ($\eta = 9.8$ has $z=0.96$). This suggests
that our lowest viscosity run ($\eta = 2.6$) and it alone, may be
approaching the inertial hydrodynamic regime, eq.\ref{inertial}; for
more evidence of this see \cite{jury}. This run covers $20,000 \le t
\le 45,000$, implying that $t^*$ (the crossover between
eqs.\ref{viscous},\ref{inertial}) is similarly large.  A less extreme
number is obtained if one quotes instead the equivalent Reynolds number
Re$^* \simeq b^2t^*$. (This relation applies because in linear
scaling, we have Re $= bl\simeq b^2 t$.) For the middle of the given
run, Re is about 20, so Re$^*$ need not be much larger than
this. Given the smallness of the apparent $b$ values (see below), the
largeness of $t^*$ follows, as does the failure to observe a clear
inertial scaling regime (eq.\ref{inertial}) in previous simulations
\cite{laradji}.

Based on these observations, we have fit our remaining 6 datasets to
the viscous hydrodynamic scaling form, eq.\ref{viscous}. In all cases
the fits are at least as convincing as those
of\cite{laradji,lebowitz}. Despite this, we {\em definitely cannot}
interpret this data (nor that of \cite{laradji,lebowitz}) as support
for a universal viscous hydrodynamic scaling, eq.\ref{viscous}.
Fig.2(a) shows fits to $f = bt$ (deviations from the data invisible on
this scale) including our own and earlier \cite{laradji,lebowitz}
datasets. Fig.2(b) shows the fitted $b$ coefficients against the mean
time $\bar t$, defined by the middle of the fitted section of each
run. Obviously, $b$ is not constant as required: it drifts {\em
systematically} toward smaller values at later times $\bar t$, a trend
representable empirically as a weak power law, $b\simeq \bar
t^{-0.2}$.

What does all this mean? Clearly, one would expect to measure $b \sim
\bar t^{-0.2}$ if in fact one had $f = c t^z$ with $z\simeq 0.8$ (see
Fig.2(a)). We are aware of no theory predicting this value of $z$, so
it would presumably have to be interpreted as an intermediate, {\em
effective} exponent arising in the crossover region between
eqs.\ref{viscous} and \ref{inertial}. Although possible, at least two
arguments counter this interpretation. Firstly, the ``crossover", if
this is indeed what we are seeing, must be exceptionally
broad. Fig.2(b) shows that a single effective exponent governs the
entire range of data shown: any ``crossover" region covers {\em four
decades} in time. The second reason to doubt this explanation is that
for all of the DPD datasets shown in Fig.2, a fit to $f = ct^z$ yields
values of $z$ that are not close to $0.8$, but close to (and usually
slightly larger than) $1.0$. Put differently, even after subtraction
of the offsets $a$, our datasets do not join up into a continous curve
on the ($l$, $t$) plot as dynamical scaling requires. This is apparent
from Fig.1, and remains true under various replottings we have tried
(such as recalculating offsets by imposing $z=0.8$ rather than $z=1$).
\begin{figure}[!ht]
\begin{center}
\leavevmode
\epsfxsize=0.35\textwidth
\epsfbox{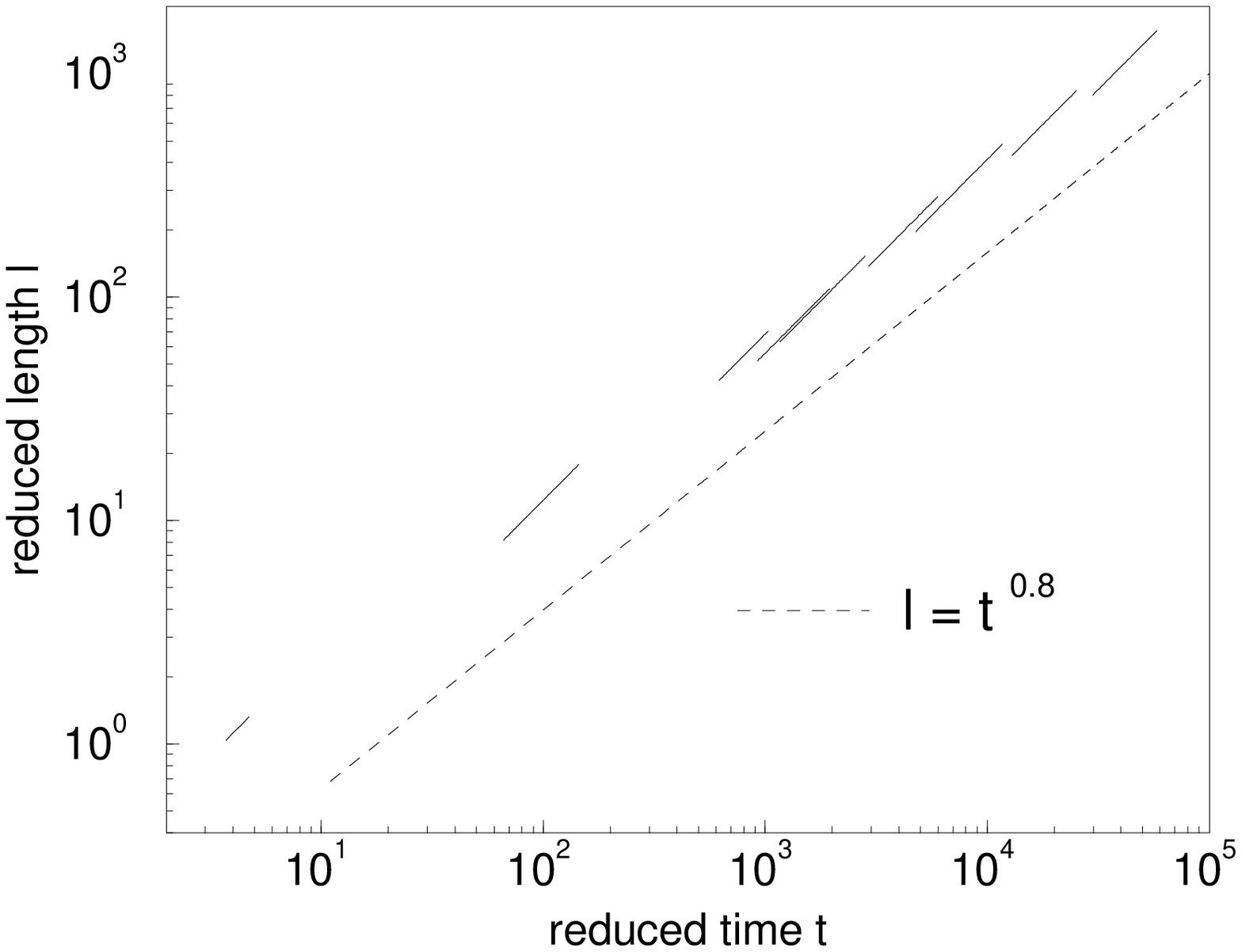}
\epsfxsize=0.35\textwidth
\epsfbox{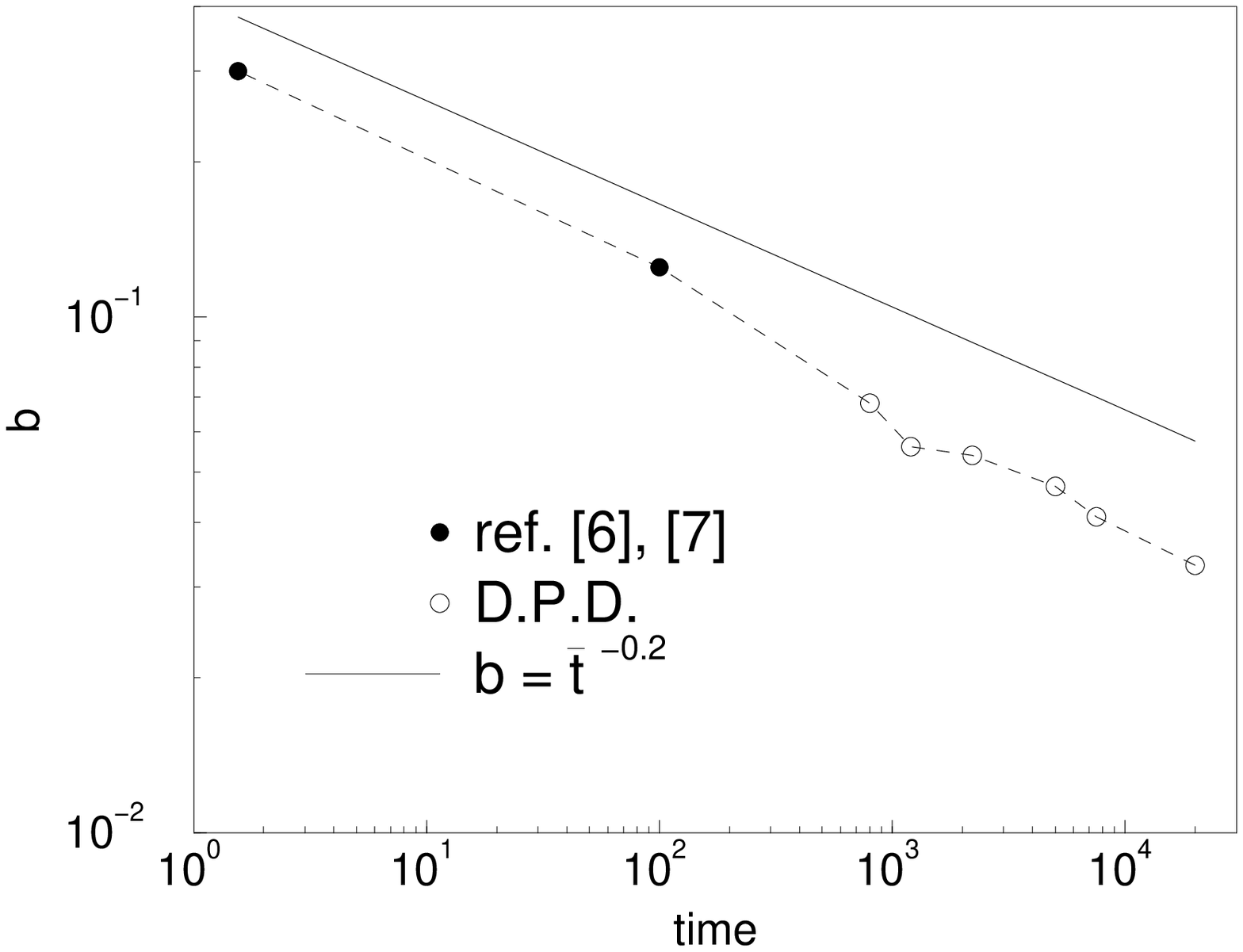}
\caption{ (a) Fitted functions $f(t) = bt$ for DPD data (rightmost 6
datasets) that of Ref. [6] (centre left) and
Ref. [7] (far left). (b) log-log plot of resulting growth
velocities $b$ against the midpoint time $\bar{t}$ of each run.}
\end{center}
\end{figure}
We therefore ask whether there might be some other physics, playing a
role in spinodal decomposition at late times, which could lead to a
violation of the dynamical scaling hypothesis itself. One possibility
is that the late-stage coarsening velocity $b$ depends on initial
conditions, inherited from the nonuniversal early-stage dynamics. (For
related ideas, see \cite{kurchan}.) This information would have to
reside either in the velocity field itself, or in subtle details of
the density distribution.  The first of these can be tested
numerically by re-initializing the fluid velocity field during a
late-stage run; we have done this and no significant effect on $b$ was
observed.

A more plausible mechanism for the observed nonuniversality of the
velocity $b$ could arise from the direct intrusion of physics that the
dynamical scaling hypothesis excludes. Thermodynamics ({\em e.g.}
finite temperature or compressibility \cite{footcomp}) cannot be
solely responsible, since all our DPD runs are {\em identical}
thermodynamically.  Perhaps the most interesting possibility is that
late-stage spinodal decomposition involves a molecular (or, in
simulations, discretization) lengthscale which could enter during
topological reconnection or ``pinch-off" events. In such events,
without which coarsening of a bicontinuous structure cannot proceed, a
fluid neck contracts to (formally) zero width in finite time.

Recent work on a closely related problem (disconnection of a single
fluid domain {\em in vacuo}) suggests that pinch-off processes need
not violate dynamical scaling \cite{eggars}: the asymptotic behavior
both before and after the pinch have a universal description in $l,t$
variables (measured from the pinch-off event itself). According to
this work, molecular physics intervenes only briefly at pinch-off, and
is forgotten soon after.  It is not yet known whether similar
universality can be recovered for fluid-fluid pinch-offs
\cite{eggars}, but crucially, even in the fluid-vacuum case, such
universality is {\em only} expected for large values of the
dimensionless quantity
\begin{equation}
\lambda = L_0 / h = \eta^2/\rho\sigma h \label{ratio}
\end{equation}
where $h$ is a molecular (or discretization) length
\cite{eggars,appert}.  For the fluid/vacuum case, Eggers \cite{eggars}
argues that $\lambda$ is large enough for some fluids ($\simeq 10^7$
for glycerol) but not others ($\simeq 20$ for water), to recover
universal behavior.

If similar ideas govern the fluid-fluid case, and if pinch-off physics
remains a controlling factor in late stage coarsening, then a
violation of dynamical scaling could be expected for many real
fluids. The same applies for any simulation in which $\lambda$ is not
very large. Taking $h=1$ (DPD units) we find that $\lambda$ in our
runs ranges from $\lambda = 0.28$ at $\eta = 12.2$ (so that $\bar t =
800$) to $\lambda = 0.014$ at $\eta = 2.6$ (so that $\bar t =
30000$). The systematic dependence of $b$ on $\bar t$ reported above
can, for these DPD runs, equally well be expressed as a dependence on
$\lambda$. The latter would permit an extended form of dynamical
scaling, with $f(t)$ replaced by $f(\lambda,t)$ in eq.\ref{dynscal};
at present this cannot be distinguished from a $\bar t$ dependence,
because the variations we make through $\eta$ affect $\bar t$ and
$\lambda$ similarly \cite{footruns}.

One speculative possibility is that the time $\Delta T$ taken for a
fluid neck, of order the domain size $L$, to reach pinch-off is not
linear in $L$ (as eq.\ref{viscous} suggests) but varies as $L
\ln(L/h)$ (so,in reduced units, $\Delta t \simeq l \ln
(l\lambda)$). In this case, individual runs would show little
departure from $f = bt$, yet $b$ would drift slowly downward with
$\bar t$, and runs of different $\eta$ would not quite superpose on
the $(l,t)$ plot. Such logarithms could conceivably arise from the
hydrodynamics of thin fluid cylinders \cite{batchelor}. A fit to $b =
\ln\eta$ (not shown) is comparable in quality, for our DPD runs, to
that in Fig.2(b) but less good than the power law shown there, if the
data of \cite{laradji,lebowitz} is included.

In conclusion, we have made a careful analysis of our extensive new
DPD data \cite{jury} and of previous simulation results
\cite{laradji,lebowitz} on spinodal decomposition in fully symmetric
binary fluids.  Taken together, these data now cover approximately
five decades in reduced physical time units. Contrary to expectation,
the data offer no clear support for the hypothesis of a universal
dynamical scaling (eq.\ref{dynscal}) \cite{julia}. Such a hypothesis
can be sustained, but only \cite{footfinite} by assuming an extremely
broad crossover between viscous ($z=1$) and inertial ($z=2/3$) scaling
regimes, with an effective exponent $z\simeq 0.8$ spanning about four
decades of reduced time $t$.  This ``slow crossover" interpretation is
more plausible when expressed in terms of Reynolds number, which spans
only $0.1 \le $Re$ \le 20$; indeed, experience with turbulence
\cite{faber} shows that a wide regime of Re might exist in which
inertial effects are significant (spoiling eq.\ref{viscous}) but not
dominant (as required for eq.\ref{inertial}).  However, such an
interpretation leaves unexplained the facts that (i) almost all our
individual simulation runs are better fit by $z\ge 1$ than $z\simeq
0.8$, and (ii) even after subtraction of nonuniversal offsets, these
runs do not lie on a common curve on the $l,t$ plot \cite{footfinite}.

We have therefore put forward a more radical proposal: that the list
of physical ingredients (fluid viscosity, density and interfacial
tension), assumed by the dynamical scaling hypothesis to dominate the
physics of spinodal decomposition at late times, is incomplete. One
possibility is that the ratio $\lambda = L_0/h$ (of continuum to
microscopic lengths) remains a relevant parameter: coarsening of a
bicontinuous structure is contingent on topological reconnection
(pinchoff) events, which could allow the intrusion of microscopic
physics no matter how large the mean domain size $L$.

We thank P. Coveney, V. Kendon, P.  Warren, and J. Yeomans for
discussions. SIJ thanks Unilever PLC and EPSRC (UK) for a CASE
Award. Work funded in part under EPSRC E7 Grand Challenge.


\begin{thebibliography}{99}
\bibitem{footlength} We convert the data of \cite{lebowitz}
to our own definition of $L$. Conversion factors are deduced from our
own simulations; this is valid so long as eq.\ref{dynscal} actually
holds.
\bibitem{footbased} These comments are based in part on more numerous
runs made for smaller system sizes.
\bibitem{footfinite} Alternatively one could assume that finite size 
effects, within each run, extend far below the length scale at which
visible saturation of the $L(T)$ curves begin. (No universal scaling
could then be expected in our data: see Fig.1.)  Our data provides no
independent evidence to support this idea; see Ref.\cite{jury}.
\bibitem{footruns} A dependence of $b$ on $\bar t$ could be
distinguished from one on $\lambda$ by doing larger simulations,
systematically varying the particle density, or by using different
simulation methods to address the same range of $t$. (The latter is
work in progress.) We can make no direct comparison with
\cite{laradji,lebowitz} since the definition of $h$ must depend on the
simulation method; but if we take $h$ as the mean interparticle
spacing, we can estimate that $\lambda \simeq 1.7$ for the data of
Ref.\cite{laradji} and $\lambda
\simeq 50$ for that of Ref.\cite{lebowitz}.
\end{thebibliography}
\end{document}